# Remote sensing of seawater and drifting ice in Svalbard fjords by compact Raman LIDAR

**Alexey F. Bunkin,[1] Vladimir K. Klinkov,[1] Vasily N. Lednev,[1] Dmitry L. Lushnikov,[1] Aleksey V. Marchenko,[2] Eugene G. Morozov,[3] Sergey M. Pershin,[1] and Renat N. Yulmetov[1,2]**

[1]*Wave Research Center, Prokhorov General Physics Institute, Russian Academy of Science, Vavilov str., 38, Moscow, Russia*

[2] *The University Centre in Svalbard, Longyearbyen, Norway*

[3] *Shirshov Institute of Oceanology, Russian Academy of Sciences, Nahimovsky ave., 36, Moscow, Russia*

[*]*Corresponding author: abunkin@kapella.gpi.ru*

A compact Raman LIDAR system for remote sensing of sea and drifting ice was developed at the Wave Research Center at the Prokhorov General Physics Institute of the RAS. The developed system is based on a diode pumped solid state $YVO_4$:Nd laser combined with compact spectrograph equipped with gated detector. The system exhibits high sensitivity and can be used for mapping or depth profiling of different parameters within many oceanographic problems. Light weight (~20 kg) and low power consumption (300 W) make possible to install the device on any vehicle including unmanned aircraft or submarine system. The Raman LIDAR presented was used for Svalbard fjords study and analysis of different influence of the open sea and glaciers on the water properties. Temperature, phytoplankton, and dissolved organic matter distributions in the seawater were studied in the Ice Fjord, Van Mijen Fjord and Rinders Fjord. Drifting ice and seawater in the Rinders Fjord were characterized by the Raman spectroscopy and fluorescence. It was found that the Paula Glacier strongly influences the water temperature and chlorophyll distributions in the Van Mijen Fjord and Rinders Fjord. Possible applications of compact LIDAR systems for express monitoring of seawater in the places with high concentration of floating ice or near cold streams in the Arctic Ocean are discussed.

## Introduction

The interest to the research in the Arctic Ocean and polar areas has been increasing in the last decades due to the growing needs for oil and gas production and to the fact that the Arctic region is a sensitive global climate indicator. Continuous diagnostics of human activity influence on the ecology and global climate changes are the most pressing goals for research in all countries located in the Arctic region. Depending on the research goals different techniques can be used for laboratory measurements while monitoring of vast areas can be carried out only by remote sensing [1-3]. Human activity influence on ecology can be traced by pollution detection: oil [4-6], chemicals or heavy metals in the ocean water [7,8]. If chemical contamination of sea is of the primary interest i.e. for ecology applications then remote detection by laser fluorescence [9] or by laser induced breakdown spectroscopy can be carried out [10-13]. Global climate change is studied on the basis of the temperature and chlorophyll distribution in the Arctic Ocean using remote methods: spaceborne radars [1,14,15], airborne microwave and laser scatterometers [16-19].

The Arctic Ocean is an excellent region for revealing climate changes and at the same time the Arctic Ocean has a strong influence on the climate. In the northern regions, the ice study is of the great interest due to the fact that ice serves as an effective interface between the ocean and the atmosphere, restricting exchange of heat, mass, momentum and chemical constituents. Detailed study of the global climate system and prediction of future climate changes can be made only if reliable data on different physical parameters of water and ice are available. These parameters include water and ice temperature, salinity, ice surface roughness, optical and thermodynamic properties of the snow cover, and the presence of liquid water above ice, which frequently occurs in summer when ice and snow melt at the ice surface.

Conventional techniques for temperature detection include spaceborne radiometer radars [1], airborne microwave and laser scatterometers. However, temperature detection by satellite radiometer is performed in a thin 30 μm surface layer of seawater. It was found that under low speed wind conditions the temperature of surface layer was by 0.5 – 1 $^{0}$C lower than in the water column [20]. The temperature detected by radars can be strongly influenced by surface waves. Frazil or grease ice can significantly change the results of measurements with such techniques in the Arctic region. Specifically, in the first stage of freezing small ice crystals are formed at the surface and as the freezing continues the crystals coagulate and form frazil ice. This frazil ice damps the short gravity waves at the sea surface, which has significant impact on radar remote sensing of the open ocean water. It was suggested to coordinate all data obtained by different techniques under one project to improve reliability of satellite data [21].

The Raman spectroscopy is a promising method for remote detection of water temperature. This method has been successfully used by different scientific groups both in the laboratory [22-25] and field experiments [27-33]. Additionally, any system for Raman spectroscopy is based on the laser source that can be used as a source of excitation for chlorophyll or organic matter fluorescence. The Raman systems for remote sensing can be used for monitoring of several parameters simultaneously: water temperature, chlorophyll and distributed organic matter (DOM), sea water contaminations by gas or oil production activity [34,35]. Two approaches can be used for remote detection of water temperature by the Raman spectra, i.e. OH-band profile fitting [25,26] and OH-band profile center correlation on temperature [22,24]. According to the first method, the Raman OH-band profile is fitted with five components attributed to different influence of surrounding medium on a water molecule [26]. Different components of OH-profile depend differently on temperature therefore intensity ratio of these components can be used to detect water temperature. The accuracy of temperature detection by the method is generally about 1 $^{0}$C [25]. The second procedure was developed by our group; it can be characterized by improved temperature accuracy detection that is better than 0.5 $^{0}$C [22]. Ice temperature can be also detected by the Raman spectroscopy but the error will be higher [18]. Raman systems can be installed on airborne platforms that would be especially useful for express diagnosis of large areas in any region including hardly accessible areas like the Arctic Ocean or mountain lakes. If gated detector is used for signal digitizing then range resolved study can be carried out including depth profiling of temperature, chlorophyll or organic mater distribution. Recent progress in laser technologies and high-speed electronics give opportunity to develop compact system for ranged Raman measurements. Small mass and compact size of such systems provide a possibility to install these devices on aircrafts or submarines including remotely driven platforms that can be extremely useful for Arctic exploration.

In this study we have carried out field experiments on laser remote sensing of drifting ice and seawater in Svalbard fjords by compact Raman LIDAR. This LIDAR is a prototype of the system that was developed for installing on unmanned aircrafts or underwater vehicles. The main goals of the research were to test the system in the field experiment conditions and to perform remote sensing of fjords: (i) to characterize floating ice in seawater; (ii) to determine temperature spatial distribution and depth profile for different types of fjords; (iii) to detect chlorophyll and organic matter in different sea environments.

## Experiment

### *Location.*

The experiment was organized by the Department of Arctic Technology of the UNIS (The University Centre in Svalbard) within the joint project of the General Physics Institute (Russia) and UNIS (Norway). The expedition took place on August 29 - September 1, 2011 in the Ice Fjord and Van Mijen Fjord and also in the Rinders Fjord close to the Paula Glacier. The LIDAR was installed on the research vessel “Viking Explorer” (length 15 m, deadweight 25 tons) owned by the UNIS. The expedition route based on the GPS data is shown in Fig. 1. The study regions included three different fjords. The Ice Fjord, Van Mijen Fjord and Rinders Fjord are good examples of the different influence of the open sea and glaciers on water mass transport and temperature distribution in Svalbard fjords. The Ice Fjord is strongly influenced by the Greenland Sea while glaciers influence is not high. The Van Mijen Fjord opening to the ocean is blocked by Akseloya Island and consequently the ocean impact is suppressed by the glaciers influence. The Rinders Fjord is an example of a strong impact of glacier and almost no influence of the sea. The fjord depth at Akseloya Island reaches 50 m while the depths in the open sea and fjord exceed 200 m and 80 m respectively. The three locations of the study are presented in Fig.

1: (1) Ice Fjord; (2) Akseloya Island and (3) Van Mijen Fjord and Rinders Fjord. All the measurements were made during the Arctic day time.

## *LIDAR.*

The LIDAR system was developed at the Wave Research Centre of the General Physics Institute of the RAS and it is presented in Fig. 2. The LIDAR is based on a compact diode pumped solid state $YVO_4$:Nd laser (Laser Compact, Ltd., model LCM-DTL-319QT: 527 nm, 5 ns, 1 kHz, 200 μJ/pulse). Beam diameter at $1/e^2$ level at the laser output is estimated at 1.8 mm, the beam divergence is 0.8 mrad. Two prisms were used to align the probe laser beam and optical path of the receiving channel. Signals from a remote object were collected by a quartz lens ($F$ = 21 cm, diameter 9 cm) and directed by an aluminum coated mirror ($l$ $x$ $h$ 4$x$3 cm) to the input slit of the spectrograph. Signal collecting efficiency was estimated in the air by elastic scattering from concrete wall at 12 m distance by spot diameter at the spectrograph slit and its value was ~250 μm. The band-pass glass filters were used to suppress scattered laser irradiation. Detection system consists of compact spectrograph (Spectra Physics, MS127i) equipped with ICCD detector (Andor iStar). We used low resolution grating 150 lines/mm and slit width 250 μm to increase signal-to-noise ratio. Spectral resolution was 120 $cm^{-1}$ that was detected by measuring of FWHM for scattered laser line. The chosen spectral window was 500 - 750 nm that made possible to register several signals simultaneously. The echo-signal is a sum of Mie, Rayleigh (527 nm), Raman (OH-bond center 650 nm) scattering, fluorescence from organic components (600 nm) and chlorophyll "a" (685 nm). The signals for elastic and Raman scattering, fluorescence were determined as the integral of the corresponding band with background correction. The ICCD detector allowed us to obtain gated images with 5 ns duration and 0.25 ns step. The time jitter between laser pulse and detection gate was less than 3 ns that provided a depth resolution of 0.85 m (the column high was 0.55 m with an error of 0.31 m) for underwater measurements. In all experiments we used a gate of 5 ns. Time - of - flight spectra measurements were carried out by varying the delay between laser pulse and exposure gate. The gated detection effectively suppressed background sun irradiance scattered by seawater. During the whole period of our experiments there was cloudy weather and there was no direct sunlight. Fast gate and the absence of direct sunlight resulted in very low level of background emission that was comparable to the noise of the detector. The system was limited to low repetition rate 23 Hz due to the maximum detector digitizing rate. The experimental setup was controlled by a PC and the software allows us to perform automatic measurements depending on the goal: sea depth profiling or mapping study. Light weight (~20 kg), small dimensions (60x40x20 cm) and low power consumption (~300 W) make it possible to install such system on any vehicle including unmanned aircraft system.

Compact Raman LIDAR system was installed in ship's bridge. Owing to the favorable weather conditions the signals were detected through an opened illuminator of the bridge. Additional aluminum coated mirror with a diameter of 30 cm was placed at a distance of 2 m from the ship board and 2 m above the seawater to guide laser beam and receiving signals to/from the studied object. The position of the sensor relative to the sea level was determined by varying delay between laser pulse and detected Raman scattering from seawater. The maximum value of the Raman signal indicates that the upper edge of the detected column was at the sea level. The measured distance between the detector and sea level was about 6 m. Laser beam was directed to the sea surface at $90^{o}$ and beam waist was placed at a distance of 2 m from ship board to prevent possible influence of wake waves.

## Results

Several points of interest for remote sensing in the Svalbard fjords were selected according to the expedition goals discussed above. A typical spectrum of ice-free seawater in the Ice Fjord is presented in Fig. 3. The spectrum (Fig. 3a) is a sum of the Mie, Rayleigh scattering, Raman scattering, fluorescence from dissolved organic material and chlorophyll. Drifting ice and sea water spectra that were detected in the Van Mijen Fjord 20 km away from Paula glacier are presented in Fig. 3. A detailed profile of OH-band and characteristic spectrum profile changes for ice and seawater can be observed in the inset in Fig.3. Fluorescence of chlorophyll "a" and organic matter were not detected in seawater in the fjord since the ice has a glacier origin and fjord seawater is efficiently mixed with the water from the glacier.

### *Laboratory experiments*

Before field experiments we have carried out laboratory experiments under controlled conditions. The first objective was to calibrate our equipment for remote temperature detection by Raman spectroscopy. The second goal was to study freezing process in detail to describe floating ice in fjords.

Previously we have proposed a method for remote temperature detecting based on Raman spectrum of water OH-band [22]. The tank with a volume of 1.5 $m^3$ was filled with the seawater from the Ice Fjord and under controlled temperature conditions Raman spectra was detected. Laser beam was directed vertically into the tank by an aluminum coated mirror placed 1 m above the water surface. The detection system was placed in another room and the spectra were collected through a glass window. The length of optical path from laser output to water surface was about 5 m. The Raman spectra were detected at different temperatures of both seawater and distilled water. According to the suggested procedure [22] every spectrum was fitted with the Gaussian profile (Fig. 4a) for different temperatures. The centers of fitted curves were plotted versus temperature for calibration of LIDAR (Fig. 4). It should be noted that the suggested procedure is very sensitive to profile variations and can be effectively used for remote temperature study. In our first paper [22] we have compared the temperature detected by our system and by a thermocouple. The differences were within a range of 0.5 $^{o}$C that was estimated as the error of this technique. In the case of compact LIDAR the results were poorer and the accuracy was about 1 $^{o}$C (Fig. 4). The slopes of curves for seawater and distilled water differed by 18%. The water salinity should be known for reliable detection of water temperature by Raman spectroscopy. The same problem arises in the detection of temperature by fitting of OH-band profile [25].

The capabilities of the Raman spectroscopy for ice thickness detection by remote sensing were demonstrated for the ice formed in seawater. The previously used tank was exposed to -8 $^{0}$C for different time periods to obtain different thickness layers of ice. The Raman spectra were obtained for a spatial column of 0.8 m for seawater covered with ice. The ice thickness was estimated by drilling holes in the ice with further measurements by a sliding meter. Ice surface was polished by melting before measurements. For several different ice thickness spectra were determined and the results are presented in Fig. 5. Ice absorption is low in the spectral range of our interest [36] (absorption coefficient is 0.28 m$^{-1}$ for wavelength equal to 650 nm) so ice absorption could be ignored for ice thickness below 0.5 m. In our experiments, the layers thickness was below 10 cm and ice can be assumed to be optically thin for the Raman spectra of seawater. In the case when the ice was only a few mm thick spectra can be described as a combination of ice and seawater spectrum. For the ice thickness greater than 1.5 cm Raman spectra were nearly the same representing characteristic spectrum for ice.

## *Field experiments in fjords*

Two types of remote sensing profiling can be used in seawater: mapping and sea depth profiling. During the measurements over the entire ship route in the Ice Fjord (region 1 in Fig. 1) we did not detect any significant changes of signals. The signals reproducibility (signal standard deviation) during the route in the Ice Fjord was 3 times greater compared to our laboratory measurements, and the amplitude deviations did not exceed 10%. These facts were explained by intense mass and heat transport of seawater from the Greenland Sea to the Ice Fjord that resulted in the uniform distribution of temperature, chlorophyll and organic material.

Different echo-signal spectra were obtained during the measurements in the Rinders Fjord closer to the Paula Glacier (region 3 Fig.1) and the corresponding spectra are shown in Fig. 6. It can be seen that the fluorescence signal from chlorophyll and DOM disappeared when the ship approached the glacier front. The Raman spectra profile also changed and became more similar to the profile that is characteristic of ice. This is explained by the strong impact of melted water from the Paula Glacier on fjord water that resulted in temperature decrease and absence of plankton.

The surface water temperature, chlorophyll and organic material were automatically mapped during the ship route back from the Paula glacier to the Rinders Fjord and Van Mijen Fjord as it is shown in Fig. 7. Data were collected automatically; each spectrum was summed from 1000 pulses during 43 s (150 m spatial averaging). The spectra were collected during 6 hours (from 11:30 till 17:55 GMT+1, August 31, 2011) with a period of 450 sec and corresponding points are marked with red squares on the map. Detected signals were normalized by the Raman signal for the following reasons. The Raman signal is proportional to excitation laser energy and concentration of water molecules. Water molecules concentration is constant and normalization by the Raman signal will correct all fluctuations of the incident laser irradiation due to ship's rocking or wind waves influence on the detected elastic signal or chlorophyll fluorescence. Additionally we have observed by the eyes that there was a high concentration of solid particles near the glacier front. These particles originate from the erosion processes and with melted water particles are transported to the fjord. When the ship was approaching the glacier front the elastic scattering enlarged due to increased concentration of dissolved solid particles. Raman intensity was decreasing since laser irradiation was effectively scattered by these solid particles. This was the second reason for normalization by the Raman signal. The elastic to Raman signals ratio decreased with increasing distance from the glacier that correlates with the decreased concentration of solid particles in water. Chlorophyll to Raman ratio was used to correct chlorophyll distribution in water. Chlorophyll was increasing while the ship was moving closer to Akseloya Island. This dependence is explained by the influence of melted water from the Paula Glacier which did not contain any organic material. Consequently, all chlorophyll in the Van Mijen Fjord was transported from the open sea while this influence was suppressed by Akseloya Island that blocked the interaction between the fjord water and the seawater from the open ocean. The temperature of the surface layer of water (thickness 0.8 m) was detected by the Raman spectra and by Conductivity, Temperature and Depth profiler (CTD) (Sea Bird, SBE-19). Since CTD measurements need a ship stop and work of the operator, the seawater parameters were determined only at the first 8 points while moving from the glacier front. The difference between temperature measurements by the Raman spectra and direct CTD results did not exceed 1 $^{0}$C. The profile of temperature was rather smooth for the surface layer of seawater at a distance greater than 20 km from the glacier front.

Additional experiments with depth profiling as well as the chlorophyll and temperature distribution were carried out near Akseloya Island to verify capabilities of compact LIDAR; the results are presented in Fig. 8. Signals were detected in a wide spectral range so all spectra detected from various depths were corrected according to absorption coefficients for different wavelengths with literature data [37].

Chlorophyll fluorescence delay should be taken into account for depth profile measurements. It is recognized in literature that chlorophyll fluorescence delay depends on different environment while delay is generally less than 4 ns [38,39]. Effective temporal resolution of compact Raman LIDAR was 8 ns, hence fluorescence delay could be assumed insignificant for the presented results. We have observed a small maximum in chlorophyll distribution at depth 2 m with slow decrease for greater altitudes.

Raman OH-band signal was detected at 5 m depth but temperature measurement procedure (described above) needs good quality spectrum with high signal-to-noise ratio. Eventually, developed compact Raman LIDAR can measure seawater temperature at depths up to 3 m only. The seawater temperature was greater for 1 $^{0}$C at the surface compared to layers at 2-3 m depth.

The widely spread echo sounding is a convention technique for safe navigation in shallow fjords. Optical ranging can be also used for depth profiling with an accuracy beyond 10 mm. We have carried out additional experiments with the developed Raman system for fjord depth profiling. When the ship entered the Van Mijen Fjord near Akseloya Island an elastic scattering of laser in the Raman system was measured for underwater depth profiling and the result is presented in Fig. 9. Elastic scattering was detected at different depths and compared with the data determined by the echo sounder. The signal detected below the fjord bottom was attributed to the surface waves that deflect the laser beam.

Compact Raman system for remote sensing of ocean is of growing interest since these systems can give reliable information about water temperature, water contaminations and chlorophyll concentration in locations that can hardly be accessed (Arctic region or mountain lakes).

One of the possible applications is to study the iceberg evolution during its motion in the open ocean. Direct measurements of iceberg parameters, i.e. temperature, salinity, mechanical properties of ice, dimensions are difficult and dangerous to perform. If a compact system will be installed on remotely operated aircraft then properties of iceberg and surrounding water can be traced automatically and safely. Laser ranging with excitation laser from the Raman system can be used to measure iceberg form and size for above and under water parts with a resolution below 5 cm. The temperature of iceberg and surrounding water can be remotely detected by Raman spectroscopy of OH-band profile that are of great importance for the study of iceberg thermodynamics. The Rayleigh, Mie and Raman scattering can reveal different ice properties as porosity, salinity, and density of ice. Additionally iceberg melting evolution and its influence on nearby seawater can be studied in detail with remotely driven LIDAR system. This study can indicate perspective procedures for express simple detection of icebergs for safe sailing in the northern seas.

Compact system can be installed on stationary platform in the Arctic Ocean as a real time automatic alarm service to detect approaching icebergs. Laser ranging can be used for mapping sea surface and automatic detection of floating objects which can be dangerous for the platform. Raman spectroscopy can be used to validate that floating object is ice. Additionally such system can be also used for express diagnostic of oil leaks at early stages in immediate vicinity of the platform.

Combination of compact Raman system and remotely operated ship or submarine can be used for automatic detection of water properties along an arbitrary path in any place of interest. These data can improve the satellite data reliability for the temperature distribution for detecting global climate changes.

## Conclusions

Remote sensing in the Svalbard fjords was carried out by compact LIDAR system. The system was installed on a small ship but can be installed on any vehicle including remotely operated aircraft since it has small weight (~ 20 kg) and low power consumption (< 300 W). The evaluation of water properties in fjords with different influence of the Arctic Ocean and Svalbard glaciers were carried out by the compact Raman system: Raman spectroscopy and fluorescence of chlorophyll and dissolved organic matter. The influence of glaciers and open sea was determined by the mapping of surface water temperature, chlorophyll and organic matter distribution. The Raman spectroscopy was used for remote temperature detection of water. The temperature profiles for depths and mapping application were detected in the Ice Fjord, Van Mijen Fjord, and Rinders Fjord. Possible applications of the compact Raman LIDAR for express monitoring of seawater properties in the places with high concentration of floating ice in the ocean are discussed.

## Acknowledgements

This work was partially supported by the Russian Foundation for Basic Research (RFBR) under projects 11-02-00034, 11-02-01202, 11-05-00448 and 11-08-00076.

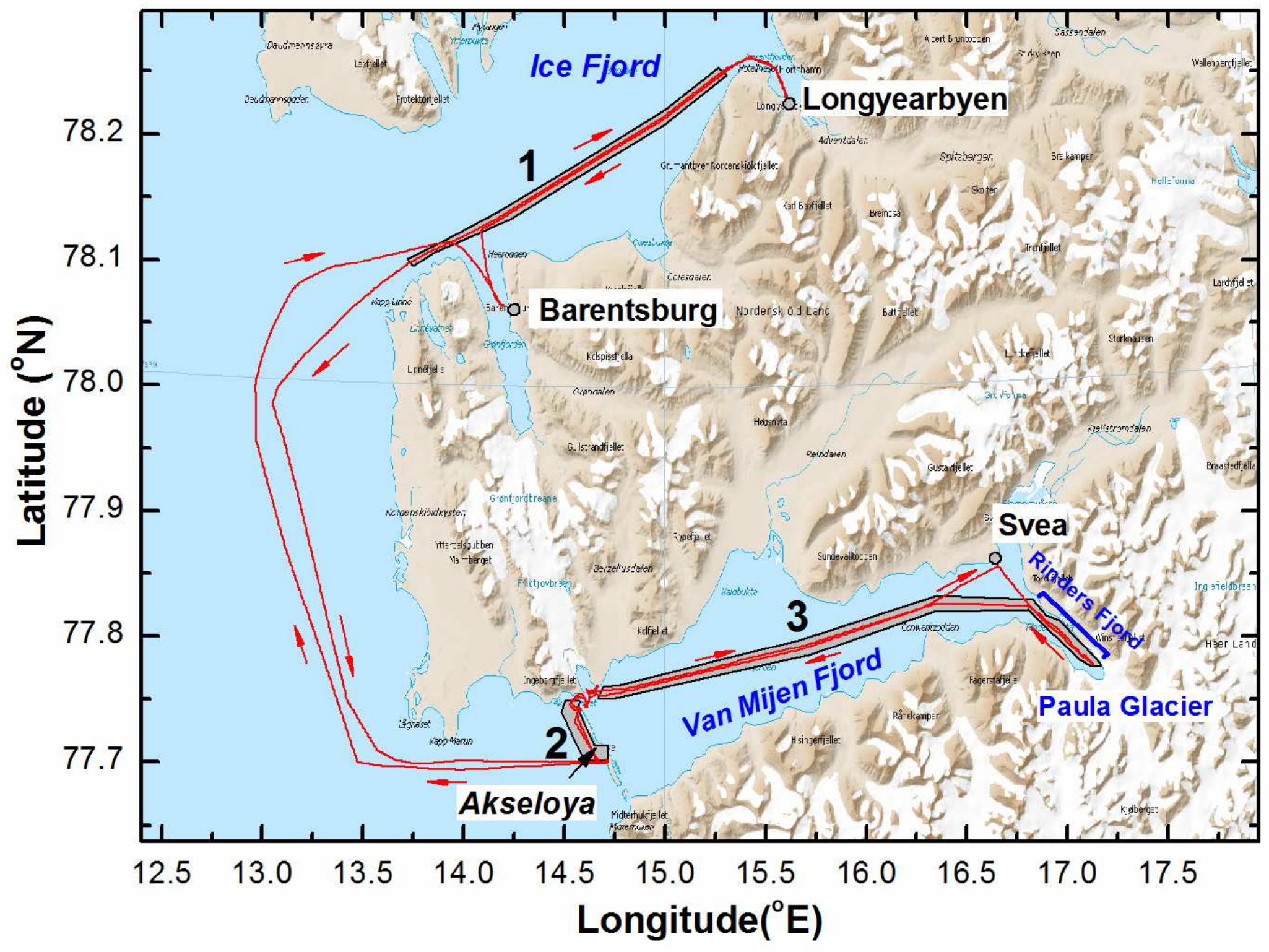

**Fig.1** Expedition map. The GPS record of the ship route (red line on map).
Remote sensing was carried out in the regions of interest that are marked with grey color:
1. remote sensing of "open" Ice Fjord; 2. near Akseloya Island at depths shallower than 20 m; 3. seawater and floating ice study in "closed" fjords, i.e. Van Mijen Fjord and Rinders Fjord.

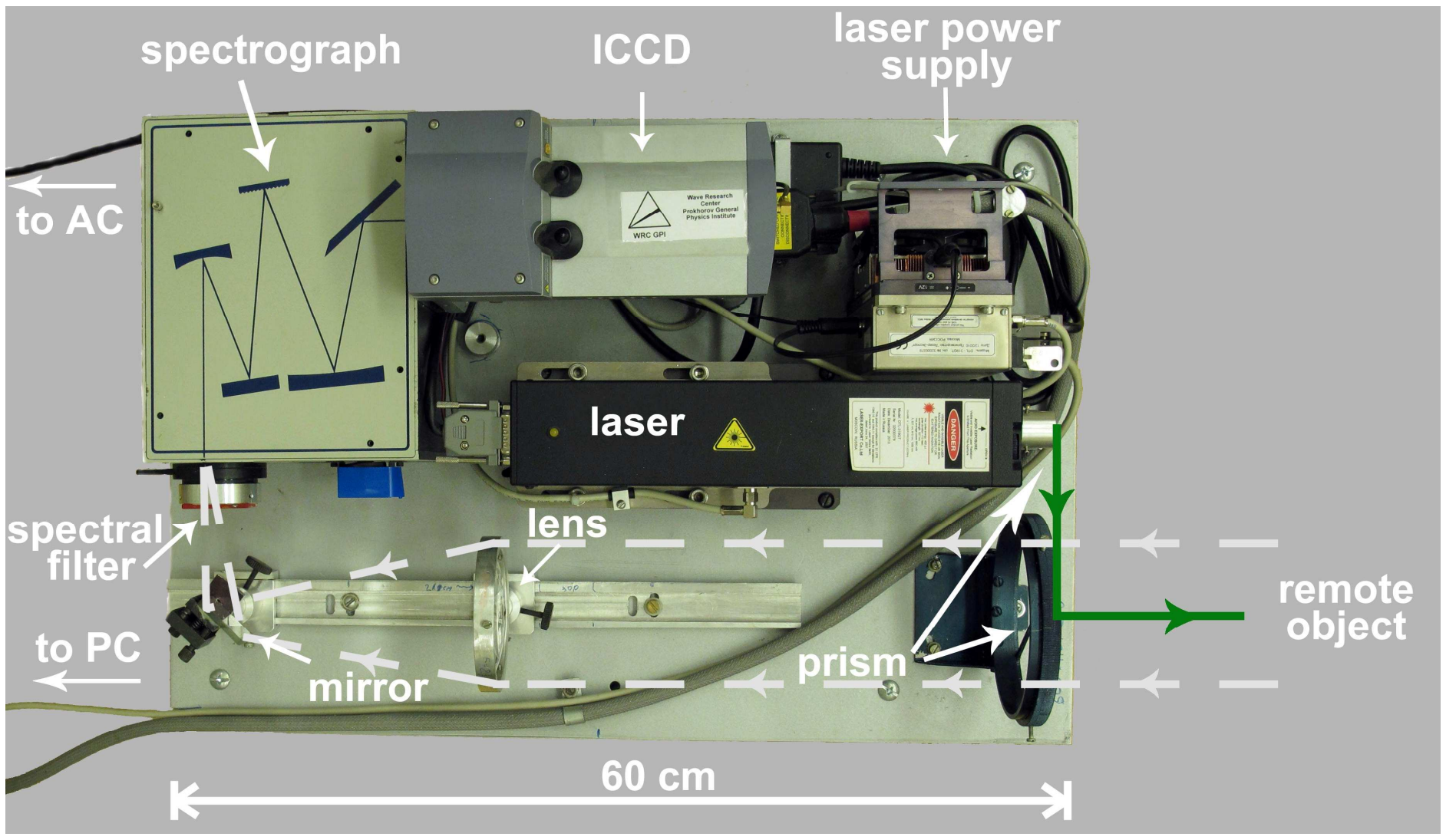

**Fig. 2** Compact Raman LIDAR system developed at WRC GPI.
Two prisms were used to guide and turn the laser beam (DPSS Nd:YAG) to remote object. Quartz lens collected the signal from the remote object and with aluminum coated mirror focused on spectrograph slit. Band pass glass filter was used to suppress laser irradiation. System dimensions were 60x40x20 cm with a mass of about 20 kg.

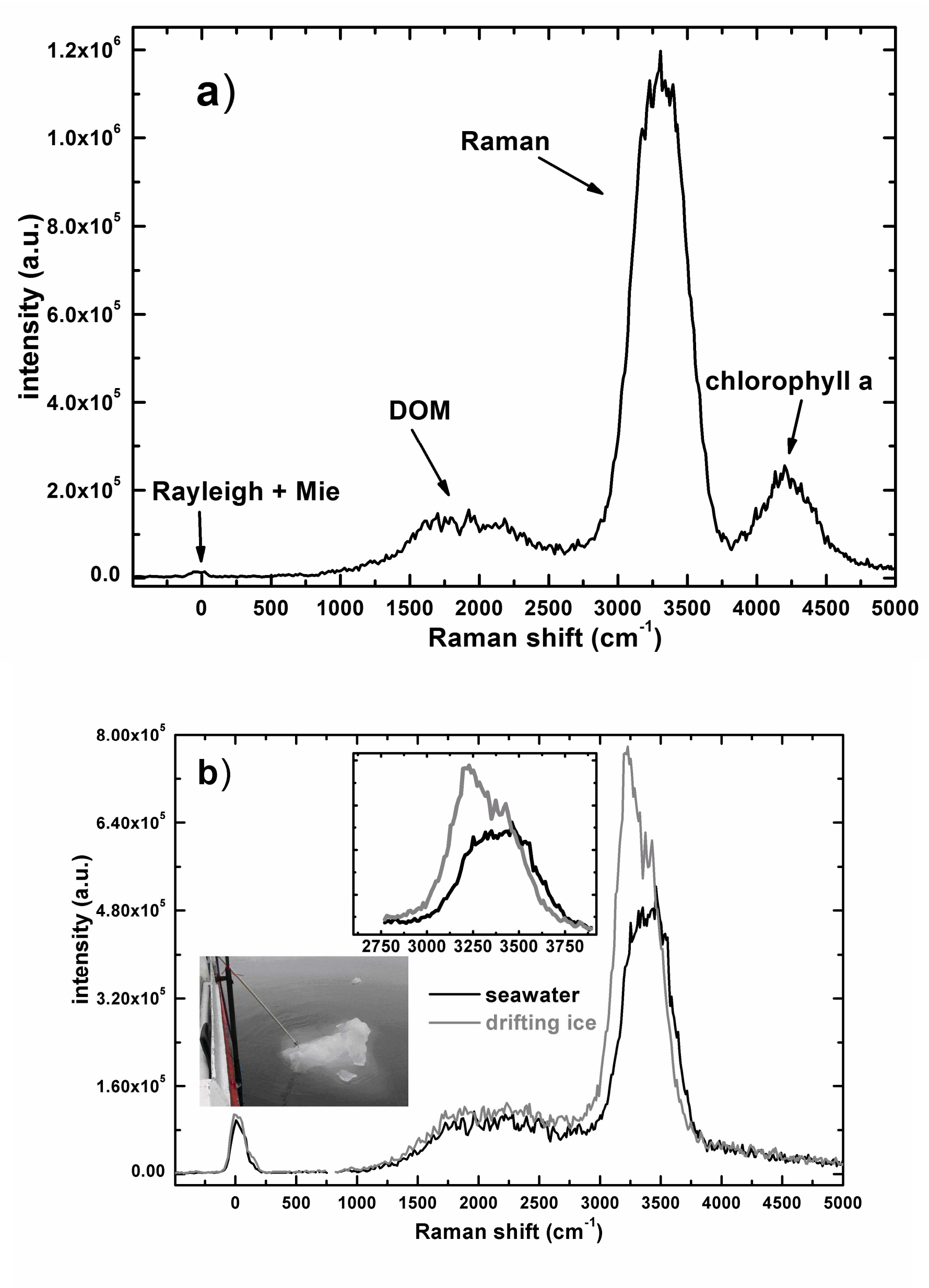

**Fig.3** Typical spectrum of seawater and floating ice.

a) Spectrum of seawater in the Ice Fjord (region 1 in Fig. 1)

Spectrum is a sum of Mie, Rayleigh scattering, Raman scattering, fluorescence from dissolved organic material (DOM) and chlorophyll

Detection was made with a delay starting 1-5 cm below the sea surface; the altitude of detection was 0.8 m (gate 5 ns); spectrum was a sum of 1000 pulses; repetition rate was 23 Hz; ship speed was 7 mph and the presented graph was spatially averaged over the intervals of 150 m

b) Spectrum of drifting ice and seawater in the Van Mijen Fjord (region 3 in Fig. 1)

In the left insert a view of floating ice and seawater is presented. Spectra were normalized by elastic scattering for a better view

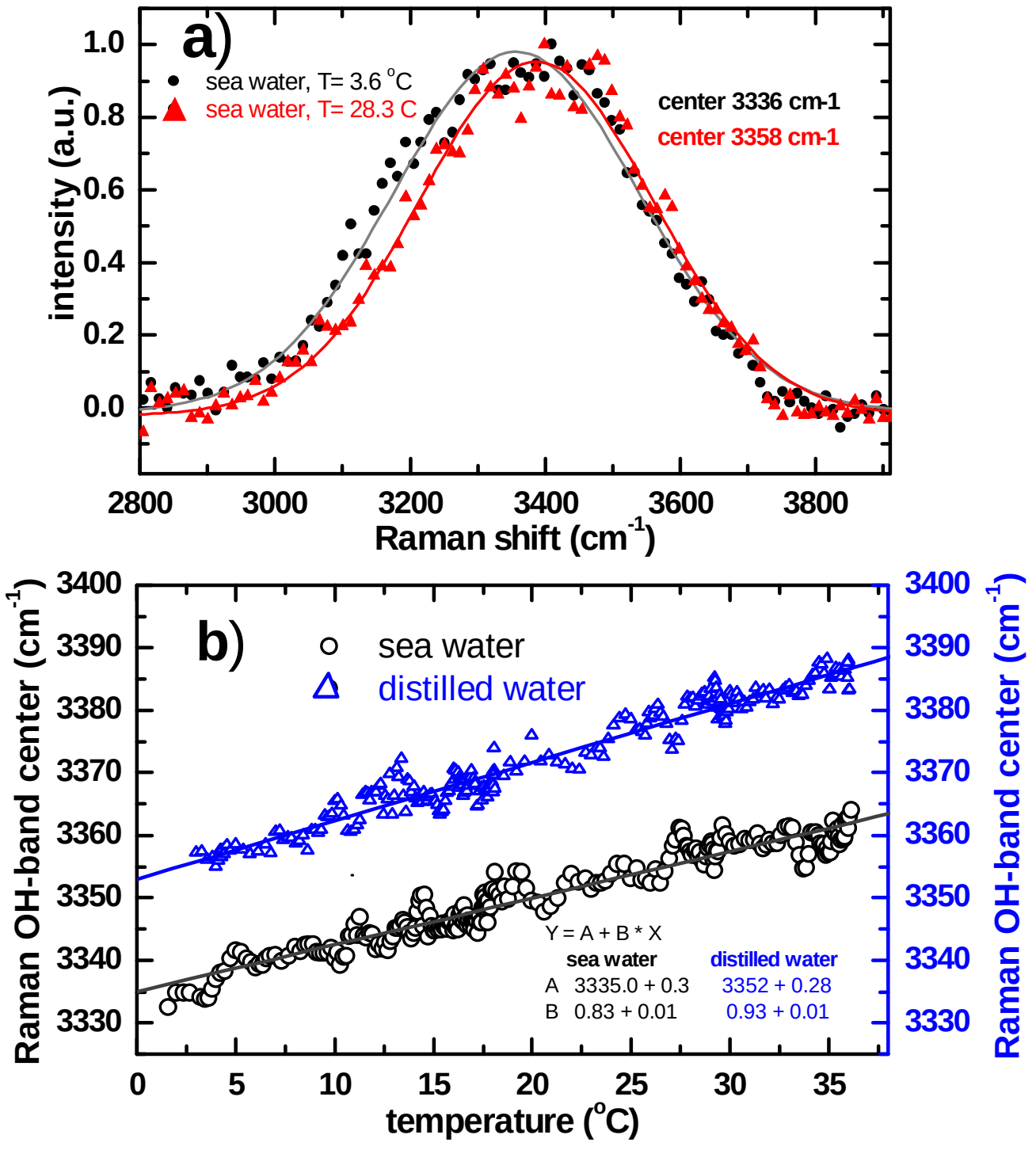

**Fig. 4** System calibration for temperature detection by Raman spectroscopy
a) Raman OH-band fitted with the Gaussian profile for seawater at different temperatures
b) Raman OH-band centers versus temperature in distilled water and seawater

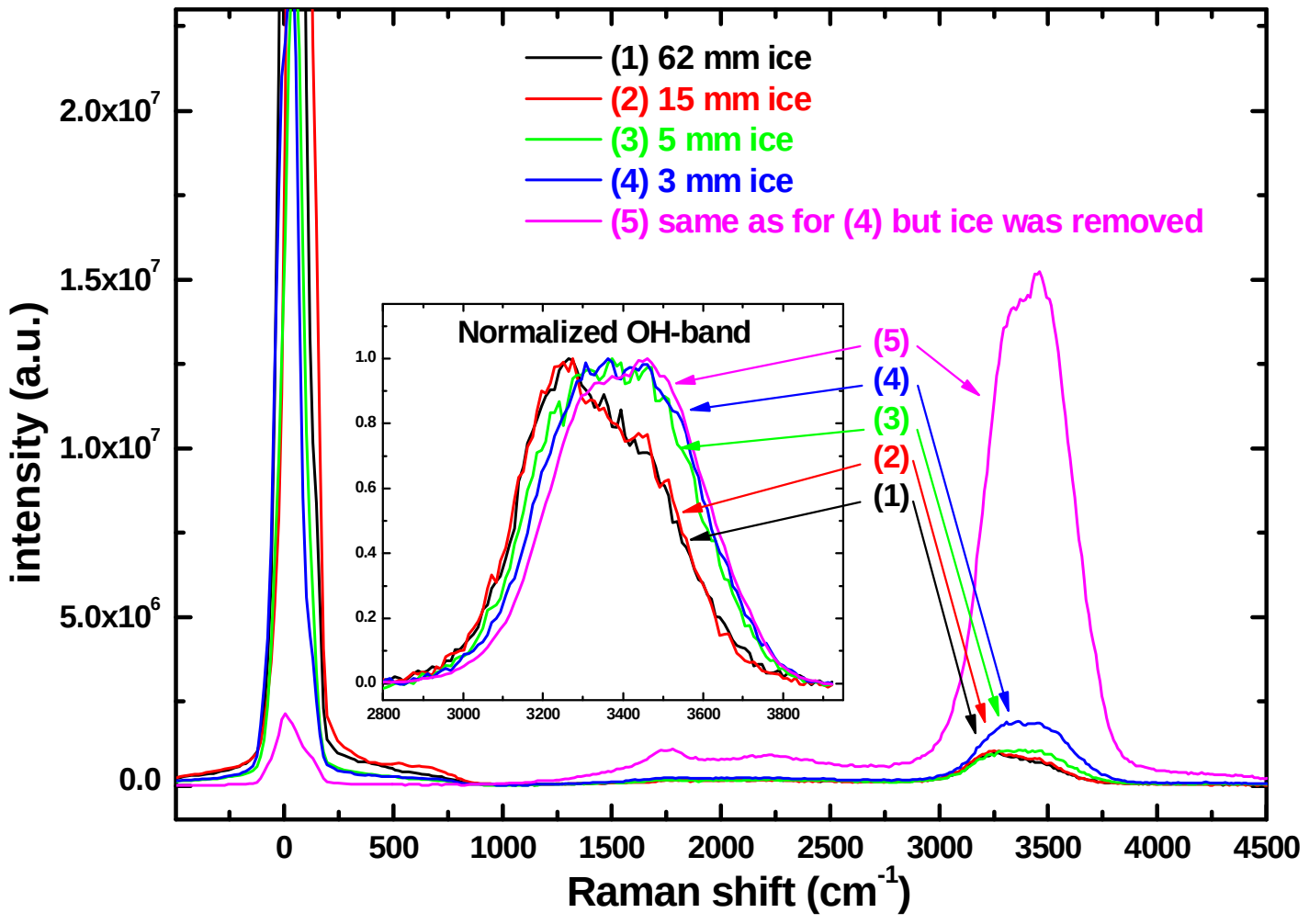

**Fig. 5** Seawater spectra without and with the ice layer of different thickness
Spectra were averaged over 1000 pulses and the detection gate (5 ns) gives spatial averaging for 0.8 m

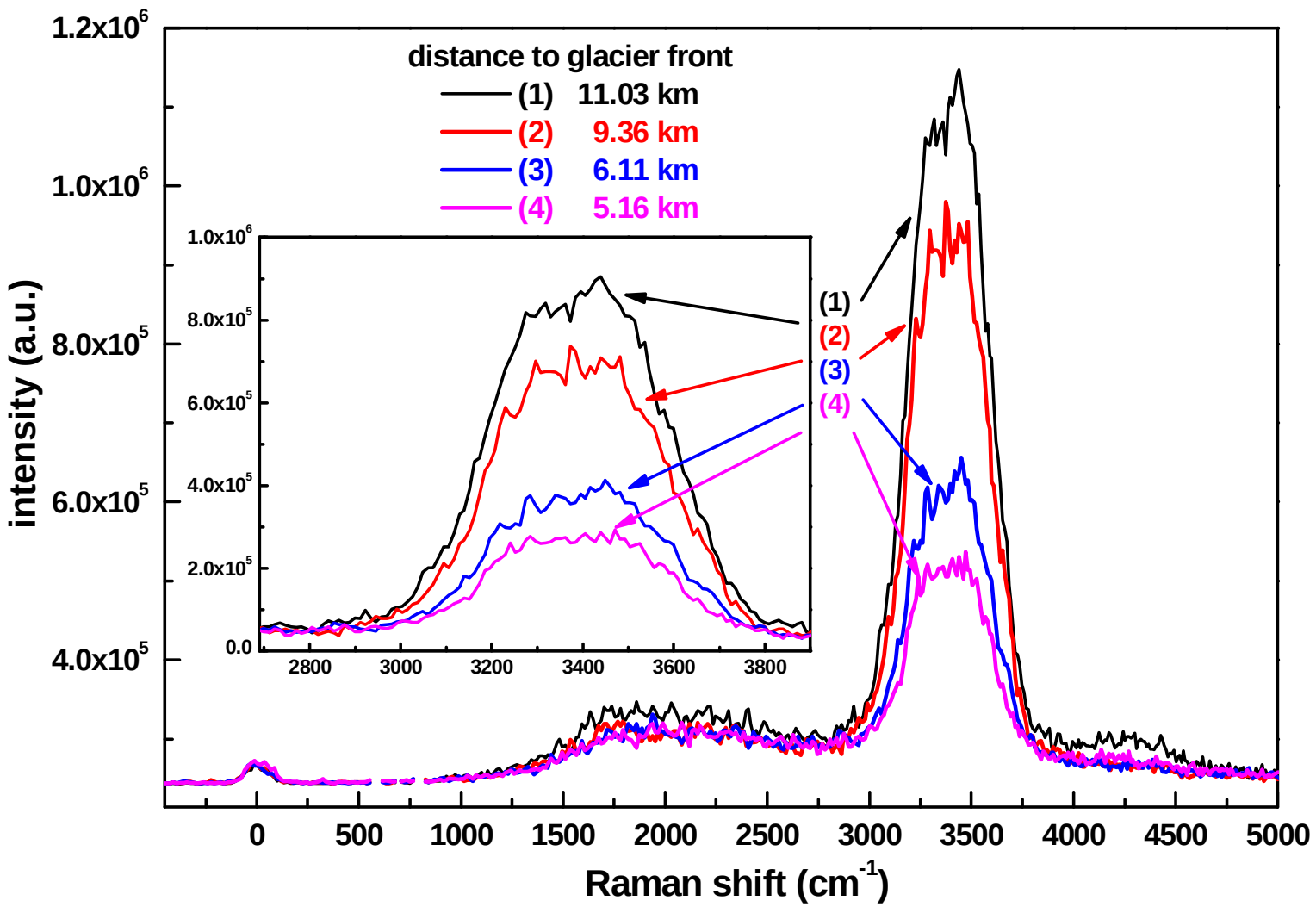

**Fig.6** Spectra for seawater in the Rinders Fjord at different distances from the Paula glacier
OH-band profile changes can be observed
Spectra were normalized by the elastic scattering signal for a better view

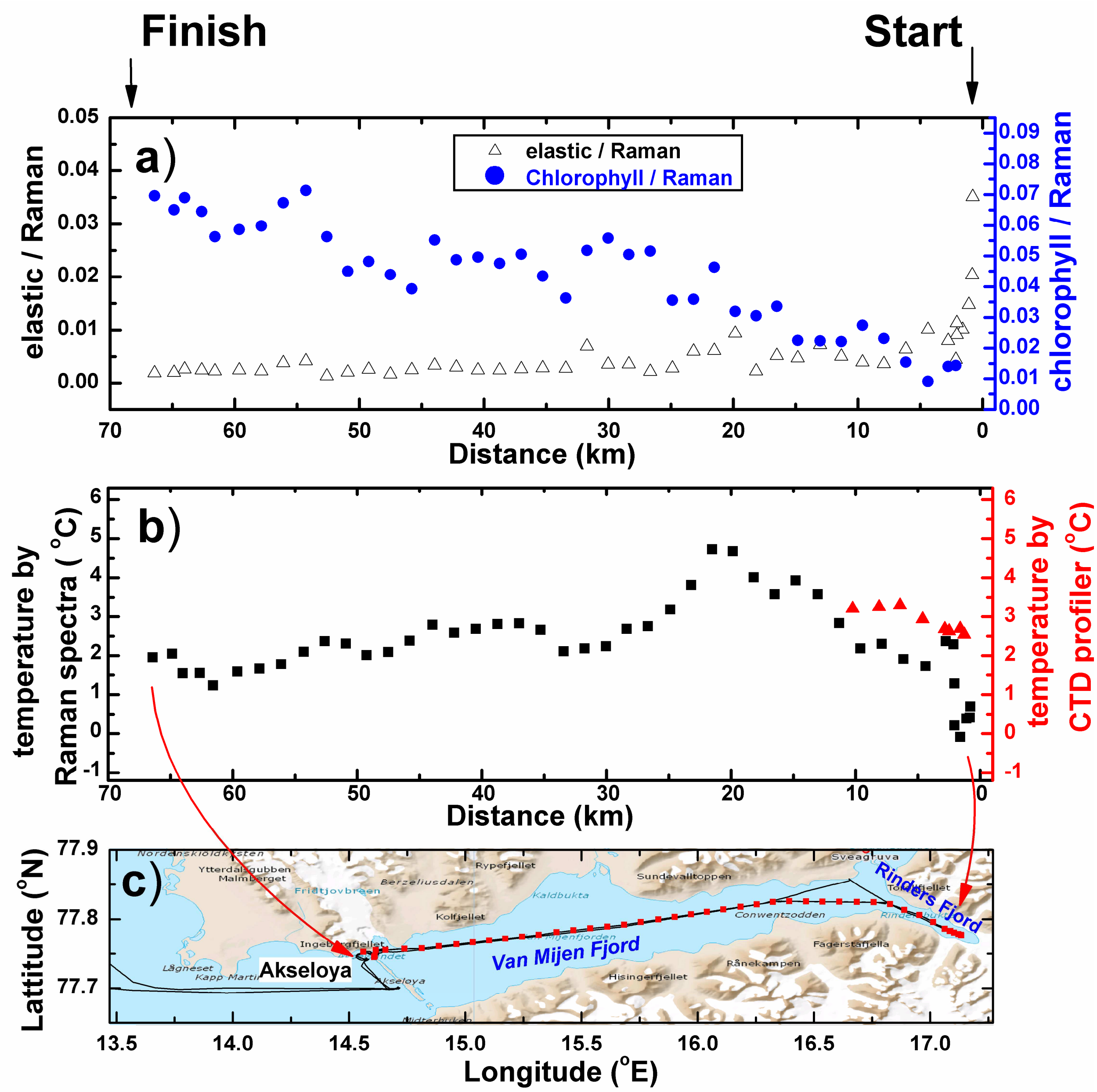

**Fig. 7** Mapping of elastic scattering (open triangles) and chlorophyll (blue circles) distribution (a), temperature (black squares) (b) in the Van Mijen Fjord and Rinders Fjord by Raman system. Temperature distribution detected by CTD-profiler is presented with red triangles (b). Expedition route and mapping points are marked with red squares (c).

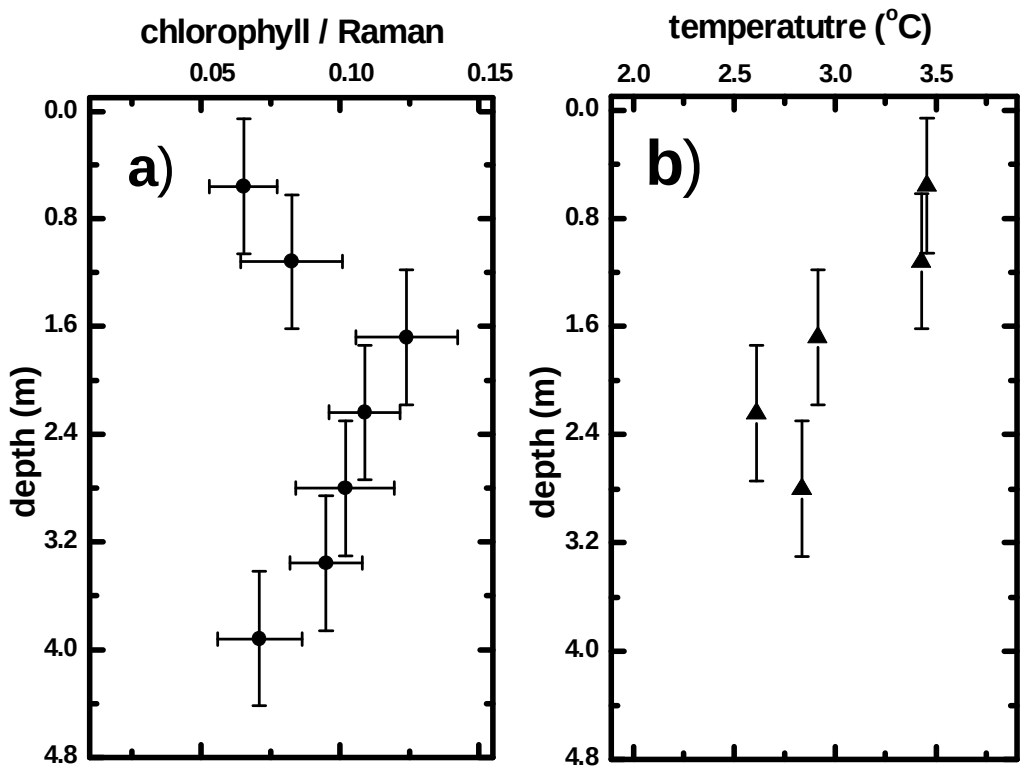

**Fig. 8** Depth profiling of chlorophyll and temperature in seawater

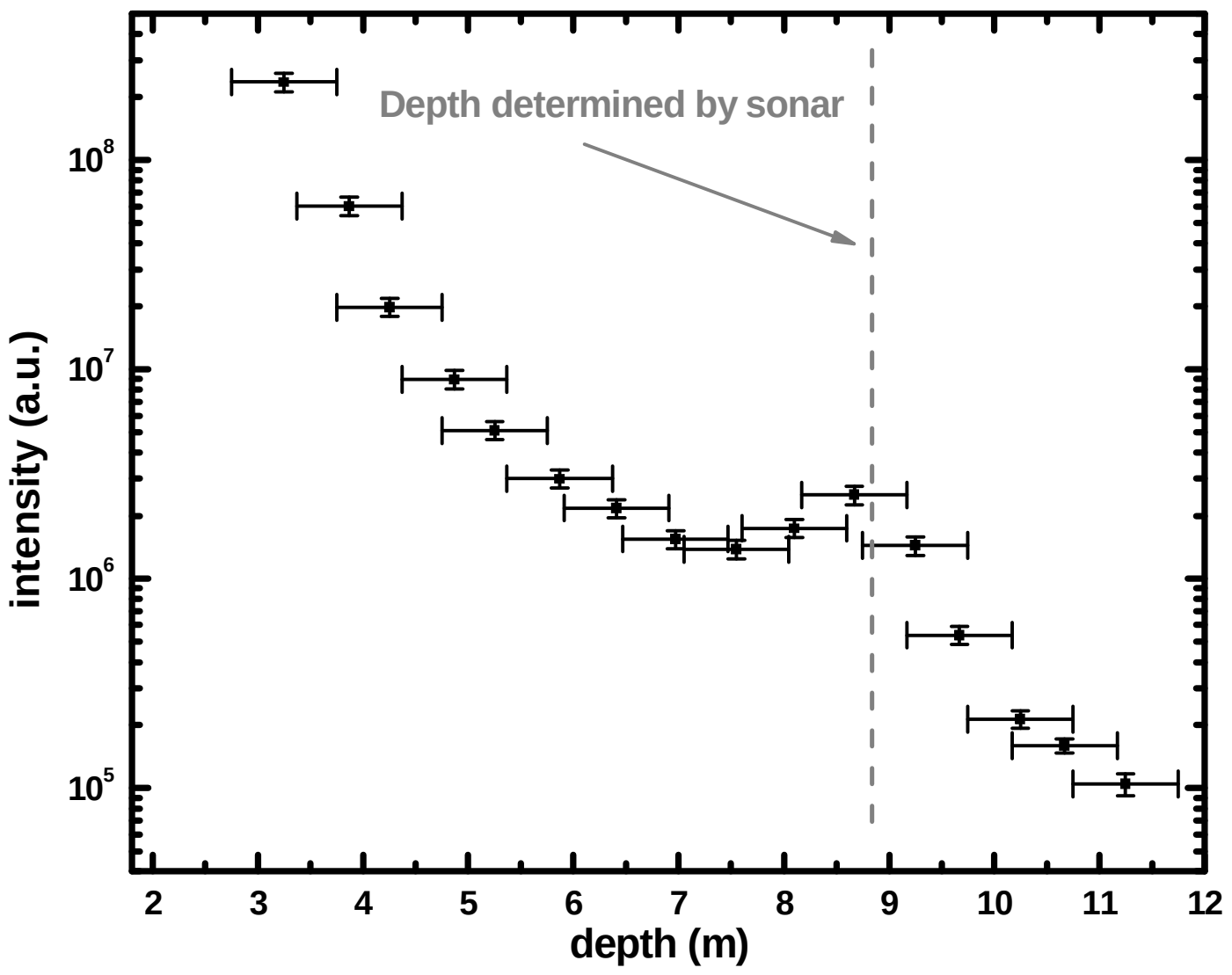

**Fig. 9** Detection of the fjord depth by elastic scattering

Additional possibility of detecting depth by the developed system was tested in field experiments

Detection accuracy of depth is in a good agreement with the echo sounder